\documentclass[twocolumn,showpacs,showkeys,prc,aps,amsmath,amssymb]{revtex4}
\usepackage{graphicx}
\usepackage{dcolumn}
\usepackage{bm}

\begin{document}

\title{Fast, Efficient Calculations of the Two-Body Matrix Elements of the Transition Operators for Neutrinoless Double Beta Decay }

\author{Andrei Neacsu}
\email{nandrei@theory.nipne.ro}
\affiliation{Horia Hulubei Foundation and Horia Hulubei National Institute for Physics and Nuclear Engineering (IFIN-HH) \\
407 Atomistilor, Magurele-Bucharest 077125, Romania}

\author{Sabin Stoica}
 \email{stoica@theory.nipne.ro}
\affiliation{Horia Hulubei Foundation and Horia Hulubei National Institute for Physics and Nuclear Engineering (IFIN-HH) \\
407 Atomistilor, Magurele-Bucharest 077125, Romania}

\author{Mihai Horoi}
 \email{mihai.horoi@cmich.edu}
\affiliation{Department of Physics, Central Michigan University, Mount Pleasant, Michigan 48859, USA}

\date{\today}

\begin{abstract}
\noindent
{\bf Background:} To extract information about the neutrino properties from the study of neutrinoless double-beta ($0\nu \beta \beta$) decay one needs a precise computation of the nuclear matrix elements (NMEs) associated with this process. Approaches based on the Shell Model (ShM) are among the nuclear structure methods used for their computation. ShM better  incorporates the nucleon correlations, but have to face the problem of the large model spaces and computational resources.\\
{\bf Purpose:} The goal is to develop a new, fast algorithm and the associated computing code for  efficient calculation of the two-body matrix elements (TBMEs) of the  $0 \nu \beta \beta$ decay transition operator, which are necessary to calculate the NMEs. This would allow us to extend the ShM  calculations for double-beta decays to larger model spaces,  of  about 9-10 major harmonic oscillator shells.\\
{\bf Method:} The improvement of our code consists in a faster calculation of the radial matrix elements. Their computation normally requires the numerical evaluation of two-dimensional integrals: one over the coordinate space and the other over the momentum space. By rearranging the expressions of the radial matrix elements, the integration over the coordinate space can be performed analytically, thus the computation reduces to sum up a small number of integrals over momentum. \\
{\bf Results:} Our results for the NMEs are in a good agreement with similar results from literature, while we find a significant reduction of the computation time for TBMEs, by a factor of about 30, as compared with our previous code that uses two-dimensional integrals.\\ 
{\bf Conclusions:}  We developed a new, fast, efficient code for computing the TBMEs that are used to calculate the NMEs necessary for the analysis of the $0 \nu \beta \beta$ decays. A rearrangement of the expressions of the radial matrix elements allow us to perform only one integration instead of computing two-dimensional integrals.  This leads to a significant reduction of the computational time, which makes us confident that it is now possible to rapidly, accurately, and efficiently calculate the TBMEs for many major harmonic oscillator shells.  
\end{abstract}

\maketitle

\section{Introduction}
The neutrinoless double beta ($0 \nu \beta \beta$) decay is a beyond Standard Model (SM) process of major interest for understanding the neutrino properties.
Indeed, its discovery would decide if neutrinos are their own antiparticles \cite{sv82}, and would give a hint on the scale of their absolute masses.
That is why there are intensive investigations on this process, both theoretical and experimental.
The present status of these investigations can be found in several more recent reviews \cite{AEE08}-\cite{VES12},
which also contain therein a comprehensive list of references in the domain.
Of particular interest is the effective neutrino mass, a parameter entering the $0\nu \beta \beta$ decay half-lives, which depends on the neutrino masses, neutrino oscillating parameters and Majorana phases.
Until now this decay mode has not yet been confirmed by independent measurements and thus, one can only extract upper limits of this parameter from the existent experimental lower limits of the half-lives.
However, to do this we also need a precise computation of the nuclear matrix elements (NMEs) which also enter the half-lives formula.
An accurate calculation of these NMEs is one of the most important challenges in the theoretical study of the $0\nu \beta \beta$ decay.

Typical calculations of the NMEs are performed using a bare transition operator \cite{VES12}.
This is almost always the case even if one uses different approaches:  pnQRPA \cite{ROD07}-\cite{SK01}, Shell model(ShM)\cite{Cau95}-\cite{HS10}, IBA \cite{BI09,iba-prl12}, 
PHFB \cite{RAH10} and energy density functional (EDS) method \cite{RMP10}-\cite{MGS11}, which are the most common methods of calculation of these matrix elements.
This is equally true even if one uses an improved transition operator that considers higher order effects in the nucleon current (HOC) \cite{sim-97,sim-09}.
In principle the most reliable of these approaches to perform calculations for the NMEs (relevant for $0 \nu \beta \beta$ decay) is the ShM, since it incorporates all types of correlations and uses effective nucleon-nucleon (NN) interactions which are checked with other spectroscopic calculations for nuclei from the same region.
However, it has to face the problem of the large model spaces and the associated computational resources.
Also, it is well known that in ShM calculations of the two-neutrino ($2 \nu \beta \beta$) matrix elements the Gamow-Teller operator needs to be quenched, to better describe the experimental data for beta decays and charge-exchange reactions.
Therefore, it is important to know if the $0 \nu \beta \beta$ transition operator has to be effectively modified when used in relatively small model spaces.
Work in this direction was recently reported in Ref. \cite{EH09} where an effective operator was analyzed for the $0 \nu \beta \beta$ decay of $^{82}Se$ in the $jj44$ model space consisting of the $f_{5/2},\ p_{3/2},\ p_{1/2},\ g_{9/2}$ orbitals. For these calculations up to 8 major harmonic oscillator shells (MHOS) were used, which implies that one needs all two body matrix elements of the $0 \nu \beta \beta$ transition operator in these large spaces. In addition, there were recent proposals \cite{medex11-mh,medex11-je}
to investigate the modifications of the transition operator in increasingly larger shell model spaces for a fictitious $0\nu\beta\beta$ decay of a $p-$ shell nucleus.

The calculations reported in Ref.\cite{EH09} were performed using a bare operator without higher order contributions in the nucleon current.
In these calculations the integral over momentum in the transition operator can be analytically done, which makes the calculation of its two-body matrix elements very fast.
It is however known that the effect of the higher order contribution in the nucleon current is a reduction of the $0\nu\beta\beta$ matrix element by 20-30 $\%$.
This reduction could be further amplified by the equivalent effective operator.
Therefore, it is important to investigate this effect, which would require knowledge of the two-body matrix elements of the bare transition operator in a large number of MHOS, e.g. 8 to 12.

In our previous works \cite{HS10}, \cite{[HSB07]},  we started to develop an efficient nuclear ShM approach to accurately calculate the NMEs for both $2\nu\beta\beta$ and $0\nu\beta\beta$ decay modes.
The approach used in Ref. \cite{HS10} to calculate the two-body matrix elements (TBME) of the
transition operator that includes higher order terms in the nucleon current needs to calculate two-dimensional integrals,  on the relative momentum and the relative coordinate.
This approach was sufficiently fast for calculating the two-body matrix elements in a single major shell, such as $pf$-shell.
However, calculations of these two-body matrix elements in 8-12 major shells would be intractable with this approach.

In this paper we present a new improved (fast, efficient) ShM code which reduces substantially the computing time of calculation of the two-body matrix elements of the transition operators for the  $0\nu\beta\beta$ decay.
In a simpler version of the code when finite nucleon size (FNS) and higher order nucleon currents correction (HOC) effects are not included, the integrals over the neutrino potentials can be performed only in the coordinate space, as has been done in Ref. \cite{Cau95}.
In the full version of the code, where these effects and the short range correlations (SRC) effects are included, normally two-dimensional integrations need to be done, one in the coordinate space and one in momentum space. 
The main improvement in this code is a rearrangement of the two-body matrix elements that allows us to do the radial integrals (the integrals in coordinate space) analytically when harmonic oscillator $\left(HO\right)$ single particle wave functions are used. Therefore, only the integration over the momentum remains to be performed numerically.
We first compare our results for NMEs with other similar results from literature performed with both ShM and other methods.
Then, we compare the CPU times of our code with the CPU times of our previous code \cite{HS10}, for the same calculations. We note that these times decrease significantly.
We get an estimation of an average CPU time per TBME and note that the new code proves very promising for more elaborate calculations in many MHOS.
The paper is organized as follows: in section 2 we describe our formalism giving the relevant expressions for the radial integrals.
In section 3 we present and discuss our results and the last section is devoted to the conclusions.
We complete the paper with an Appendix where expressions of several quantities used in the formalism described in section 2 are given.

\section{Description of the formalism}
The $0\nu \beta \beta$ decay $(Z,A)\rightarrow (Z+2,A)+2e^{-}$ requires that the neutrino and the antineutrino are identical and massive particles. Considering that this decay occurs  only by exchange of light neutrinos between nucleons and in the presence of left-handed weak interactions, the lifetime can be expressed as:
\begin{equation}
\left( T^{0\nu}_{1/2} \right)^{-1}=G^{0\nu}(E_0, Z)\mid M^{0\nu}\mid^2 \left( \frac{ \left< m_{\nu} \right> }{m_e} \right)^2 \ ,
\end{equation}
$G^{0\nu}$ is the phase space factor depending on the energy decay $E_0$ and nuclear charge $Z$, and $\left<m_\nu\right>$ is the effective neutrino mass parameter depending on the first row elements of the neutrino mixing matrix $U_{ei}$, Majorana phases $e^{i\alpha_i}$ and the absolute neutrino mass eigenstates $m_i$ (see e.g. Ref. \cite{VES12}).
The nuclear matrix elements are:
\begin{equation}
 M^{0 \nu}=M^{0 \nu}_{GT}-\left( \frac{g_V}{g_A} \right)^2 \cdot M^{0 \nu}_F \ ,
\end{equation}
where $M^{0 \nu}_{GT}$ and  $M^{0 \nu}_F$ are the Gamow-Teller ($GT$) and the Fermi($F$) parts, respectively. Usually a tensor part appears as well, but the numerical calculations have shown that its contribution is small \cite{HS10}; consequently, it will be neglected in the following. The $GT$ and $F$ parts are defined as follows:
\begin{equation}
M_\alpha^{0\nu} = \sum_{m,n} \left< 0^+_f\| \tau_{-m} \tau_{-n}O^\alpha_{mn}\|0^+_i \right> \ ,
\end{equation}
where $O^\alpha_{mn}$ are transition operators ($\alpha=GT,F$) and the summation is over all the nucleon states.

Due to the two-body nature of the transition operator, the matrix elements are reduced to sum of products of two-body transition densities (TBTD)  and matrix elements for two-particle states (TBME),
\begin{eqnarray}
\nonumber M_\alpha^{0\nu} & = & \sum_{j_p j_{p^\prime} j_n j_{n^\prime} J_\pi} TBTD \left( j_p j_{p^\prime} , j_n j_{n^\prime} ; J_\pi \right) \\
& & \left< j_p j_{p^\prime}; J_\pi \| \tau_{-1} \tau_{-2}O^\alpha_{12} \| j_n j_{n^\prime} ; J_\pi \right> ,
\end{eqnarray}
The two-body transition operators $O^{\alpha}_{12}$ can be expressed in a factorized form as:
\begin{eqnarray}
O^{\alpha}_{12} = N_{\alpha}  S_{\alpha}^{(k)} \cdot  R_{\alpha}^{(k)}
\end{eqnarray}
where $N_\alpha$ is a numerical factor including the coupling constants, and  $S_\alpha^{(k)}$ and  $R_\alpha^{(k)}$  are operators acting on the spin and  relative wave functions of two-particle states.
Thus, the calculation of the matrix elements of these operators can be decomposed into products of reduced matrix elements within the two subspaces \cite{HS10}. The expressions of the two-body transition operators are:
\begin{eqnarray}
O_{12}^{GT} = \sigma _1 \cdot \sigma _2 H(r) \ , \ \ \
O_{12}^{F} = H(r) \ .
\end{eqnarray}
The most difficult is the computation of the radial part of the two-body transition operators, 
which contains the neutrino potential.  We will refer to its computation in more detail.
Neutrino potential is of Coulomb type, depending weakly on the intermediate states, and is defined by integrals of momentum carried by the virtual neutrino exchanged between the two nucleons \cite{sim-09}
\begin{eqnarray}
\nonumber H_{\alpha} (r) & = & \frac{2R}{\pi}  \int^\infty_0 j_0 (qr) \frac{h_{\alpha}(q)}{\omega} \frac{1}{\omega + \left<E\right>}q^2 dq \\
& \equiv & \int^\infty_0 j_0 (qr) V_{\alpha}(q) q^2 dq \ ,
\label{n_potential}
\end{eqnarray}
where $R=1.2 A^{1/3}$ fm, $\omega = \sqrt{q^2+m_\nu^2}$ is the neutrino energy and $j_0(qr)$ is the spherical Bessel function.
We use the closure approximation in our calculations, and $\left<E\right>$ represents the average excitation energy of the states in the intermediate odd-odd nucleus, that contribute to the decay. The expressions of $h_{\alpha} (\alpha = F, GT)$ are
\begin{equation}
h_F = G_V^2(q^2) 
\end{equation}
and 
\begin{equation}
\begin{split}
 h_{GT}(q^2) = & \frac {G^{2}_A(q^2)}{g^{2}_A} \left [ 1- \frac{2}{3}\frac{q^2}{q^2+m^2_\pi} + \frac{1}{3}\left( \frac{q^2}{q^2+m^2_\pi} \right )^2 \right ]\\
             + & \frac{2}{3} \frac {G^{2}_M(q^2)}{g^{2}_A}\frac{q^2}{4m^2_p} \ ,
\end{split}
\label{hgt-hoc}
\end{equation}
 where $m_\pi$ is the pion mass, $m_p$ is the proton mass and 
\begin{equation}
G_M(q^2) = (\mu_p - \mu_n)G_V(q^2),
\end{equation}
with $(\mu_p - \mu_n)=4.71$.

The expression (\ref{hgt-hoc}) includes finite nucleon size (FNS) and higher order terms in the nucleon currents (HOC). 
The $G_V$ and $G_A$ form factors which takes into account the FNS effects are: 
\begin{equation}
G_A \left(q^2 \right) = g_A \left( \frac{\Lambda^2_A}{\Lambda^2_A+q^2} \right)^2, \ G_V \left( q^2 \right) = g_V \left( \frac{\Lambda^2_V}{\Lambda^2_V+q^2} \right)^2
\label{formfactors}
\end{equation}
For the vector and axial coupling constants we used $g_V = 1$ and $g_A = 1.25$, 
and the values for the vector and axial vectors form factors we used are $\Lambda_V=850 MeV$ and $\Lambda_A=1086 MeV$ \cite{AEE08}.

For computing the radial matrix elements \\ $\left<nl|H_{\alpha}|n'l'\right>$ we use   
the harmonic oscillator $HO$ wave functions  $\psi_{nl}(lr)$ and $\psi_{n^\prime l^\prime}(r)$ corrected by a factor \\ $[1 + f(r)]$, which takes into account the short range correlations induced by the nuclear interaction:
\begin{equation}
\psi_{nl}(r) \rightarrow \left[ 1+f(r) \right] \psi_{nl}(r) \ .
\end{equation}
For the correlation function we take the functional form
\begin{eqnarray}
f(r) = - c \cdot e^{-ar^2} \left( 1-br^2 \right) \ ,
\label{src}
\end{eqnarray}
where $a$, $b$ and $c$ are constants which have particular values for in different parameterizations \cite{ccm}-\cite{ccm1}, as  it will be discussed in the next section. \\

As an observation, we note that in the limit $m_\nu=0 \rightarrow \omega = q$ and when HOC and FNS effects are neglected, the integral over $q$ in Eq. (7) can be easily done and the expression of the neutrino potential becomes
\begin{eqnarray}
H_{\alpha}® \proptoß \frac{1}{r}\left[\sin(\left<E\right>r) \ Ci(\left<E\right>r) - \cos(\left<E\right>r) \ Si(\left<E\right>r) \right] \ \
\end{eqnarray}
where $Si(z)$ and $Ci(z)$ are the sine and cosine integral functions \cite{sintegral}.
Then, 
the calculation of the radial integrals over the neutrino potential reduces to a single integral in the coordinate space, which can give a hint about the magnitude of the NMES within an error of about 25-35$\%$. This approximation was used in Ref. \cite{Cau95} where the NMEs for $^{48}Ca$ are calculated.  

Including HOC and FNS effects the radial matrix elements of the neutrino potentials become:
\begin{eqnarray}
\nonumber \left\langle nl \mid H_\alpha(r) \mid n^\prime l^\prime \right\rangle & = &\int^\infty_0 r^2 dr \psi_{nl}(r) \psi_{n^\prime l^\prime}(r)\left[ 1+f(r) \right]^2 \\
& \times & \int^\infty_0 q^2 dq V_{\alpha} (q) j_0 (qr) 
\ ,
\label{H-two-integrals}
\end{eqnarray}
where $\nu$ is the oscillator constant.

As one can see, if all the nuclear effects are included, the calculation of the radial integrals (\ref{H-two-integrals}) requires the numerical computation of two integrals, one over the coordinate space and the other over the momentum space. However, one can reduce the computation to only one integral proceeding as follows: one can rearrange the expression of the radial integral in coordinate space as a sum of terms with the same power of $r$. The dependence of $r$ appears from the product of the $HO$ wave functions, the correlation function and Bessel functions. First, one can write the product of two $HO$ wave functions as a sum over the terms with the same power in $r$:
\begin{eqnarray}
\nonumber \psi_{nl}(r) \psi_{n^\prime l^\prime}(r) & = & \sum_{s=0}^{n+n^{\prime}} A_{l+l^{\prime}+2s}(nl,n^{\prime}l^{\prime}) \left( \frac{2}{\pi} \right)^{\frac{1}{2}} \\
& \times & (2 \nu )^\frac{l+l^{\prime}+2s+3}{2} e^{- \nu r^2} r^{l+l^{\prime}+2s},
\end{eqnarray}
where $A_{l+l^{\prime}+2s}$ are coefficients independent of $r$ whose expressions are given in Appendix.
Then, one adds the contribution of the factor $\left[ 1+f(r) \right]^2$ that brings dependence of $r$ in powers of $0$, $2$ and $4$:
\begin{eqnarray}
\nonumber \left[ 1+f(r) \right]^2 & = & 1 -2 c e^{-ar^2} +2 b ce^{-ar^2}r^2 + c^2e^{-2ar^2} \\
& - & 2 bc^2 e^{-2ar^2}r^2 + b^2 c^2e^{-2ar^2}r^4
\end{eqnarray}
Finally, the computation of  the radial matrix elements requires to perform integrals of the form:
\begin{eqnarray}
&& \nonumber \mathcal{I}_\alpha(\mu;m)  = \int_0 ^\infty q^2 dq \ V_\alpha(q)
\\
&&\times \left( \frac{2}{\pi} \right)^{\frac{1}{2}} \left( 2 \nu \right) ^{\frac{m+1}{2}} \int_0 ^\infty dr \ e^{-\mu r^2}r^m j_0(qr)
\end{eqnarray}
where $\mu$ = $\nu$, $\nu+a$, $\nu+2a$ and  $m$ is integer. In this expression the integration over $r$ can be done analytically and one gets:
\begin{equation}
\begin{split}
& \left( \frac{2}{\pi} \right)^{\frac{1}{2}} \left( 2 \nu \right) ^{\frac{m+1}{2}} \int_0 ^\infty dr \ e^{-\mu r^2}r^m j_0(qr) = \left( \frac{2\nu}{2\mu} \right) ^{\frac{m+1}{2}} \\
&\times \left( m-1 \right)!! \sum_{k=0}^{\frac{m}{2}-1} (-1)^k {{\frac{m}{2}-1} \choose {k}} \frac{e^{-\frac{q^2}{4\mu}} }{(2k+1)!!(2\mu)^k}q^{2k}
\end{split}
\end{equation}
Thus, $\mathcal{I}_{\alpha}(\mu;m)$ becomes:
\begin{eqnarray}
\nonumber \mathcal{I}_{\alpha}(\mu;m) & = & \left( \frac{2 \nu}{2\mu} \right) ^{\frac{m+1}{2}} (m-1)!! \\
& \times & \sum_{k=0}^{\frac{m}{2}-1}(-1)^{k} {\frac{m}{2}-1 \choose {k}} \mathcal{J}_{\alpha}(\mu;k)
\end{eqnarray}
where $\mathcal{J}_{\alpha}(\mu;k)$ are integrals over  momentum:
\begin{equation}
\begin{split}
\mathcal{J}_{\alpha}(\mu;k)  = & \frac{1}{(2k+1)!!} \frac{1}{(2\mu)^k} \\
\times & \int_0 ^\infty exp \left(- \frac{q^2}{4\mu} \right) q^{2k +2}V_{\alpha}(q)
dq
\end{split}
\label{qintegral}
\end{equation}
Finally, the radial matrix element can be expressed as a sum of integrals over the  momentum space:
\begin{eqnarray}
\left\langle nl\mid H_{\alpha}(r)\mid n^{\prime} l^{\prime} \right\rangle = \sum_{s=0}^{n+n^{\prime}} A_{l+l^{\prime}+2s}(nl,n^{\prime}l^{\prime}) \mathcal{K}_{\alpha}(m)
\label{finaleq}
\end{eqnarray}
where $\mathcal{K}_{\alpha}(m)$ is a sum of six $\mathcal{I}_{\alpha}(\mu;m)$ integrals over momentum. Its expression is given in Appendix.

\section{Numerical results and discussions}

We developed a new code for computing the TBMEs necessary for the ShM calculations of the NME involved in $0\nu\beta\beta$ decays, based on the formalism described in section 2 that reduces the two-dimensional integrals in Eq. (\ref{H-two-integrals}) to sums of one-dimensional integrals of Eq. (\ref{qintegral}). The new code has a flexible user interface, which allows the selection of various  nuclear effects.
As it was shown in the previous section, the improvement consists in a simplification in the computation of the radial matrix elements which allows us to perform the integration only in the momentum space.
We first check our code comparing our results for the total NMEs, for $^{48}Ca$ and $^{82}Se$ with similar results from Refs. \cite{Cau95} and \cite{KOR07}.
In order to obtain the $M^{0\nu}$ nuclear matrix element, the two-body transition densities are calculated using the method described in Ref. \cite{HS10}. For $^{48}Ca$ we used GXPF1A \cite{gx1} effective interaction in the full $pf$ model space, and for $^{82}Se$ we used JUN-45 \cite{jun45} effective interactions in the $jj44$ model space.
 
As one can see from Table I, our results are in good agreement with previous ones, provided that the same nuclear nuclear effects are included in the calculations.
Our code enables us to include in a flexible manner the nuclear effects, such as SRC of Jastrow \cite{jastrow} type with Miller-Spencer and coupled cluster model (CCM) \cite{ccm} with Argonne V18 and CD-Bonn \cite{ccm}-\cite{ccm1} parameterizations, FNS, and HOC. The SRC parameters entering Eq. (\ref{src}) are the same as in Table II of Ref. \cite{HS10}. For comparison, we used in our calculation the same nuclear effects as those used in the corresponding references.  

\begin{table}[h]
	\begin{center}
        \begin{tabular}{|l|c|c|}
            \hline
            $M^{0 \nu}$ & $^{48}Ca$ &  $^{82}Se$ \\
            \hline
            $^{(*)}$ \  present work & 0.573 & 2.47 \\
	    \cite{HS10} \ (2010 ShM)& 0.57 \ & \  \\
	    \cite{prl100} \ (2008 ISM) & 0.59 & 2.11  \\
	    \cite{npa818} \ (2009 ISM) & 0.61 & 2.30 \\
	    \cite{su-prc75-2007} (2007 QRPA) & \ & 2.77 \\
	    \hline
        \end{tabular}
	\label{tab:comparison}
	\caption{Comparison between the results of the present work $^{(*)}$ and other similar results from the references indicated. In the calculation we used SRC of Jastrow type, FNS and HOC.}
	\end{center}
\end{table}

\begin{figure}[h]
\includegraphics[width=0.4\textwidth]{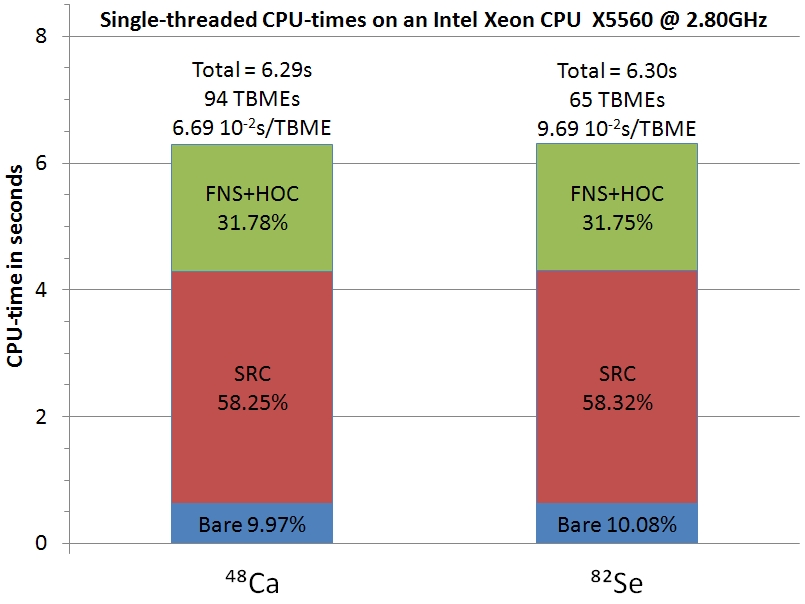}
\caption{CPU-times for the computation of the TBMEs}
\label{fig:cputimes}
\end{figure}

We have also analyzed the performance of our code in getting an improved computing speed.
In Figure \ref{fig:cputimes} we show the single-core CPU times needed to compute the TBMEs.

In the case of $^{48}Ca$, there are 94 TBMEs requiring a total of $6.29s$ of CPU-time on our test machine equipped with Intel Xeon X5560 CPUs.
This translates into an average of $6.7\cdot 10^{-2}s$ for each individual TBME.
When computing the product of wave functions, the dependence on the $n$ and $l$ quantum numbers of the nucleon orbits is reflected in the CPU-times, as one can see in the difference between the average CPU-time/TBME of $^{48}Ca$ and those of $^{82}Se$. $^{82}Se$ has required a total of $6.30s$ for the computation of its 65 TBMEs, thus needing an average time of $9.7\cdot 10^{-2}s$ for each TBME.
Even then, we can still calculate TBMEs for $^{82}Se$ almost as fast as for the simpler case of $^{48}Ca$ nucleus.

Figure \ref{fig:cputimes} also shows the contribution of the SRC and FNS+HOC effects to the total computation time for the TBMEs. "Bare" means that neither SRC nor FNS + HOC effects were considered. 
With our new method and code, we obtain an improvement in speed by a factor of about 30, as compared to the code used in Ref. \cite{HS10}, where more than three minutes where needed instead of 6.3 seconds.

The performance of the new code makes us confident that it is now possible to rapidly, accurately and  efficiently compute TBMEs for many nuclear shells. This task is very challenging for the TBME code of Ref. \cite{HS10}. For example, if one wants to investigate the effective transition operator in only 8 MHOS \cite{EH09}, one needs to calculate about 434k TBMEs ($GT$ plus $F$).
The actual average time/TBME is about 1.7 seconds, but as remarked above, is increasing with the raise of the angular momenta of the single particle orbits involved. Using a conservative estimate of about 10 seconds/TBME one could conclude that one needs about 50 days of single-threaded processing power to calculate all necessary TBMEs. This time could be reduced by a factor of say 500 if the calculation of the TBMEs is distributed via a load-balancing algorithm \cite{SeH10}, when using 1000 cores with 50$\%$ efficiency.
However, this reduction might not be sufficient if 9 or 10 MHOS need to be used. The new algorithm presented here could be extremely useful in reducing the calculation time by another factor of about 30.

\section{Conclusions}
We developed a fast, efficient code for computing the TBMEs, which are part of the NMEs necessary for the analysis of the $0\nu\beta\beta$ decays. The improvement consists in a faster computation of the  radial matrix elements using correlated $HO$ wave functions.
Their computation normally requires the numerical evaluation of two-dimensional integrals, one over the coordinate space and the other over the momentum space.
By rearranging the expressions of the radial matrix elements, the radial integrals can be performed analytically over the coordinate space, thus the computation reduces to sum up a small number of integrals over momentum.
We check our code by comparing the values of the NMEs for $^{48}Ca$ and $^{82}Se$ calculated with our new code with similar results from literature and we found a quite good agreement.
Further, we estimated the CPU-times for one single core needed to compute the TBMEs with our code and compare them with the similar CPU-times obtained with our previous code requiring two-dimensional integrals.
We find a significant reduction of the computational time, by a actor of about 30.
We also estimated the average CPU-time per single TBME in the cases $^{48}Ca$ and $^{82}Se$ and found very small values.
This achievement makes us confident that it is now possible to rapidly, accurately and  efficiently compute TBMEs for many major harmonic oscillator shells, which were very time-consuming in our earlier approach. The calculation of the TBMEs in 8 MHOS could be done in about 1-2 days using the present single-threaded code. Extension to more than 8 MHOS would require the parallelization of the code using a load-balancing algorithm.
These TBMEs can be further used to investigate the effective transition operator needed for $0\nu\beta\beta$ decay analyses. 

\section{Appendix}
\noindent
{\bf A1}. The $HO$ radial wave functions are given by:
\begin{equation}
\psi _{nl}(r) = N_{nl} \ exp \left(- \frac{\nu r^2}{2} \right) r^l \ L_{n}^{ \left( l+ \frac{1}{2} \right) }\nu r^2 \ ,
\end{equation}
\\
\text where $\nu$ is the oscillator constant, $N_{nl}$ is the normalization constant
\begin{equation}
N_{nl} = \left[ \frac{2^n n!}{\left(2l +2n +1 \right)!!} \right]^{\frac{1}{2}} \left(2\nu \right)^{\frac{2l+3}{4}} \left( \frac{2}{\pi}\right) ^\frac{1}{4}
\end{equation}
and $L_{n}^{ \left( l+ \frac{1}{2} \right) }(\nu r^2)$ is the Laguerre associated polynomials:
\begin{equation}
\begin {split}
& L_{n}^{ \left( l+ \frac{1}{2} \right) }(\nu r^2) = \frac{\left(2l +2n +1 \right)!!}{2^n n!} \\
& \times \sum_{k=0}^{n}{n \choose {k}} \frac{1}{\left(2l +2k +1 \right)!!}\left( -2\nu r^2 \right)^k \ .
\end {split}
\end{equation}
\\

\noindent
{\bf A2}. The expression for the $A_{l+l^{\prime}+2s}(nl,n^{\prime} l^{\prime})$ used in Eq. (\ref{finaleq}) is:
\begin{equation}
\begin {split}
A_{l+l^{\prime}+2s}(nl,n^{\prime} l^{\prime}) & = \left[ \frac{n!(2l+2n+1)!!}{2^n} \frac{n^{\prime}!(2l^{\prime}+2n^{\prime}+1)!!}{2^{n^{\prime}}} \right] \\
& \times  (-1)^s \sum_k \frac{1}{k! (n-k)!(2l+2k+1)!!}\\
& \times  \frac{1}{k^\prime ! (n^\prime-k^\prime)!(2l^\prime+2k^\prime+1)!!} \ ,
\end {split}
\end{equation}
with  $max(0,s-n^{\prime}) \leq k \leq min(n,s)\ , \ \ \ k + k^\prime = s$ \ . \\

\noindent
{\bf A3}. The expression for the $\mathcal{K}_{\alpha}(m)$ used in Eq. (\ref{finaleq}) is:
\begin{eqnarray}
\nonumber \mathcal{K}_{\alpha}(m) & = & \frac{1}{\sqrt{2\nu}} [ \mathcal{I}_{\alpha}(\nu ; m) - 2c\mathcal{I}_{\alpha}(\nu + a;m) \\
\nonumber &+&2c \left( \frac{b}{2\nu} \right) \mathcal{I}_{\alpha}(\nu +a;m+2) + c^2\mathcal{I}_{\alpha}(\nu +2a;m) \\
\nonumber &-&2c^2 \left( \frac{b}{2\nu} \right) \mathcal{I}_{\alpha}(\nu +2a;m+2) \\
&+& c^2\left( \frac{b}{2\nu} \right)^2 \mathcal{I}_{\alpha}(\nu +2a;m+4) ]  \ , 
\end{eqnarray}

\noindent
where $a$, $b$, and $c$ are the SRC parameters entering Eq. (\ref{src}).

\begin{acknowledgments}
Support from project IDEI-PCE Nr. 58/28.10/2011 is acknowledged.
MH acknowledges support from the USA NSF grant PHY-1068217.
\end{acknowledgments}

\end{document}